 \documentclass[final,3p,times,sort&compress]{elsarticle}

\usepackage{amssymb}
\usepackage{amsmath}
\usepackage{hyperref}
\usepackage{graphicx}
\usepackage{subcaption}
\usepackage{soul,color,xcolor}

\journal{Nuclear Physics B}

\begin{document}

\begin{frontmatter}

\title{Hawking radiation of magnetized particles via tunneling of Bardeen black hole}

\author{Baoyu Tan\corref{cor1}}
\ead{2022201126@buct.edu.cn}

\address{College of Mathematics and Physics, Beijing University of Chemical Technology,
	15 Beisanhuandonglu Street, Beijing, 100029, China}
\cortext[cor1]{Corresponding author}

\begin{abstract}
So far, no one has studied regular black holes using the Parikh-Wilczek method. In this paper, we calculated the emission rate of magnetized particles passing through the event horizon of the Bardeen black hole by using the Parikh-Wilczek method. The emission spectrum deviates from the pure thermal spectrum, but conforms to the unitary principle of quantum mechanics. Our results support the conservation of information.
\end{abstract}


\end{frontmatter}

\section{Introduction}
\label{sec:1}
In 1975, Hawking proposed Hawking radiation, which suggested that particles could escape from a black hole through quantum effects~\cite{1}. In 1976, Hawking showed that this pure thermal radiation carries no information~\cite{2}. As the black hole persists in emitting this pure thermal radiation, the information contained within the black hole will be lost. The loss of information violates a fundamental principle of quantum mechanics: the principle of unitarity. In recent years, significant progress has been made in the conservation of black hole information using the entanglement entropy method~\cite{3,4,5,6,7,8,9}. However, due to the lack of a complete quantum gravity theory, the specific form of the density matrix is difficult to obtain.

In 2000, Parikh and Wilczek identified Hawking radiation as a quantum tunneling process~\cite{10}. Their calculation results are consistent with the unitary principle and support the conservation of information. The researchers then studied various static and stationary rotating black holes through the method and came to the same conclusion that Hawking radiation deviates from the pure thermal spectrum, satisfying the unitary principle and supporting the conservation of information~\cite{11,12,13,14,15,16,17,18,19,20,21,22,23,24,25,26,27,28,29,30,31,32,33}.

But until now, no one has investigated whether the tunneling process of magnetic particles in Bardeen black holes satisfies the principle of unitarity, and whether information is conserved. In 1968, Bardeen solved a black hole without a singularity~\cite{34}. In 2000, Ayon-Beato and Garcia discovered that Bardeen black holes could be used as exact solutions to Einstein's field equations~\cite{35}. The metric of the Bardeen black hole can be given by \cite{36}:
\begin{equation}
	\mathrm{d}s^2=-f(r)\mathrm{d}t^2+\frac{1}{f(r)}\mathrm{d}r^2+r^2\mathrm{d}\Omega^2.
\end{equation}
Where:
\begin{equation}
	f(r)=1-\frac{2Mr^2}{(r^2+Q_g^2)^{3/2}}.
\end{equation}
Where $Q_g$ is magnetic charge of the black hole. It is particularly noteworthy that for regular black holes, $M$ no longer represents internal energy, and the first law of black hole thermodynamics also needs to be revised~\cite{37,38,39,40,41,42}.

In the section~\ref{sec:2}, we calculate the modified emission spectrum using the Parikh-Wilczek method. Finally, in section~\ref{sec:3}, we briefly discuss our results. We have adopted the system of natural units in this paper ($G\equiv \hbar \equiv c\equiv 1$).
\section{Tunneling rate}
\label{sec:2}
In order to eliminate the coordinate singularity, we give the Painlev\'{e} line element of Bardeen black hole:
\begin{align}
	\mathrm{d}s^2&=-\left[1-\frac{2Mr^2}{(r^2+Q_g^2)^{3/2}}\right]\mathrm{d}t^2+2\sqrt{\frac{2Mr^2}{(r^2+Q_g^2)^{3/2}}}\mathrm{d}t\mathrm{d}r+\mathrm{d}r^2+r^2\mathrm{d}\Omega^2\nonumber\\
	&=\tilde{g}_{00}\mathrm{d}t^2+2\tilde{g}_{01}\mathrm{d}t\mathrm{d}r+\mathrm{d}r^2+r^2\mathrm{d}\Omega^2.\label{eq:1}
\end{align}
For the mass particles with magnetic charges, we treat them as de Broglie s-waves and obtain the time-like geodesic equation:
\begin{equation}
	\dot{r}=v_p=\frac{1}{2}v_g=-\frac{1}{2}\frac{\tilde{g}_{00}}{\tilde{g}_{01}}=\frac{1}{2r}\frac{(r^2+Q_g^2)^{3/2}-2Mr^2}{\sqrt{2M(r^2+Q_g^2)^{3/2}}}=\frac{f(r)}{2r}\sqrt{\frac{(r^2+Q_g^2)^{3/2}}{2M}}.\label{eq:2}
\end{equation}
Where $\dot{r}$ is the phase velocity of de Broglie s-waves. The solution $r_h$ for $f(r)=0$ is the event horizon of the black hole. In order to get the radiation spectrum right, we have to take into account the effect of self-gravitation. $M$ and $Q_g$ in Eq.~\ref{eq:1} and Eq.~\ref{eq:2} should be replaced by $M-\omega$ and $Q_g-q_g$.

The matter-gravity system we consider is composed of the black hole and the electromagnetic field outside the black hole. The Lagrangian of this system is:
\begin{equation}
	L=L_m+L_g=L_m-\frac{1}{4}\tilde{F}_{\mu\nu}\tilde{F}^{\mu\nu}.
\end{equation}
The generalized coordinate is $\tilde{A}_\mu=(\tilde{A}_t,0,0,0)$. The expression for $L_g$ shows that $\tilde{A}_t$ is a cyclic coordinate. To eliminate the degree of freedom of $\tilde{A}_t$, the action should be written as:
\begin{equation}
	S=\int_{t_i}^{t_f}(L-P_{\tilde{A}_t}\dot{\tilde{A}}_t)\mathrm{d}t.
\end{equation}
According to the WKB approximation, the emission rate of tunneling particles can be written as follows:
\begin{equation}
	\Gamma\sim e^{-2\mathrm{Im}S}.\label{eq:3}
\end{equation}
The imaginary part of the action in Eq.~\ref{eq:3} can be written as:
\begin{equation}
	\mathrm{Im}S=\mathrm{Im}\left\{\int_{r_i}^{r_f}\left[P_r-\frac{P_{\tilde{A}_t}\dot{\tilde{A}}_t}{\dot{r}}\right]\mathrm{d}r\right\}=\mathrm{Im}\left\{\int_{r_i}^{r_f}\left[\int_{(0,0)}^{(P_r,P_{\tilde{A}_t})}\mathrm{d}P^\prime_r-\frac{\dot{\tilde{A}}_t}{\dot{r}}\mathrm{d}P^\prime_{\tilde{A}_t}\right]\mathrm{d}r\right\}.\label{eq:4}
\end{equation}
Where $P_{\tilde{A}_t}$ is the conjugate momentum of $\tilde{A}_t$.

In order to proceed with the calculation, we wrote down the Hamilton equation:
\begin{align}
	\dot{r}=&\frac{\mathrm{d}H}{\mathrm{d}P_r}\bigg|_{(r;\tilde{A}_t,P_{\tilde{A}_t})},\label{eq:5}\\
	\dot{\tilde{A}}_t=&\frac{\mathrm{d}H}{\mathrm{d}P_{\tilde{A}_t}}\bigg|_{(\tilde{A}_t;r,P_{r})}.\label{eq:6}
\end{align}
Put Eq.~\ref{eq:5} and Eq.~\ref{eq:6} into Eq.~\ref{eq:4}, we can get:
\begin{equation}
	\mathrm{Im}S=\mathrm{Im}\left\{\int_{r_i}^{r_f}\left[\int_{(M,E_{Q_g})}^{(M-\omega,E_{Q_g-q_g})}\frac{1}{\dot{r}}(\mathrm{d}H)_{r;\tilde{A}_t,P_{\tilde{A}_t}}-\frac{1}{\dot{r}}(\mathrm{d}H)_{\tilde{A}_t;r,P_{r}}\right]\mathrm{d}r\right\}.\label{eq:7}
\end{equation}
Where $E_{Q_g}$ is the energy of the electromagnetic field. We note that there is a difference of a factor $\mathcal{W}=1+\int_{r_h}^{+\infty}4\pi r^2\frac{\partial T^0_0}{\partial M}\mathrm{d}r$ between $M$ and internal energy, The energy changes caused by mass reduction and magnetic charge reduction can be written as:
\begin{align}
	(\mathrm{d}H)_{r;\tilde{A}_t,P_{\tilde{A}_t}}=&\mathcal{W}\mathrm{d}(M-\omega)=\mathcal{W}\mathrm{d}M^\prime,\label{eq:8}\\
	(\mathrm{d}H)_{\tilde{A}_t;r,P_{r}}=&\frac{Q_g-q_g}{r}\mathrm{d}Q^\prime_g.\label{eq:9}
\end{align}
Eq.~\ref{eq:8} and Eq.~\ref{eq:9} represent the effect of the loss of mass and magnetic charge on the energy of the black hole, respectively. Put Eq.~\ref{eq:2}, Eq.~\ref{eq:8} and Eq.~\ref{eq:9} into Eq.~\ref{eq:7}, we can get:
\begin{equation}
	\mathrm{Im}S=\mathrm{Im}\left\{\int_{r_i}^{r_f}\int_{(M,E_{Q_g})}^{(M-\omega,E_{Q_g-q_g})}\frac{2r}{f(r)}\sqrt{\frac{2(M-\omega)}{[r^2+(Q_g-q_g)^2]^{3/2}}}\left[\mathcal{W}\mathrm{d}M^\prime-\frac{(Q_g-q_g)}{r}\mathrm{d}Q^\prime_g\right]\mathrm{d}r\right\}.
\end{equation}

In order to proceed with the calculation, we switch the order of integration and first integrate $r$. Apparently, there is a pole at the event horizon $r=r_h$ of the black hole. We choose a new integral loop and apply the residue theorem:
\begin{equation}
	\mathrm{Im}S=\int_{(M,E_{Q_g})}^{(M-\omega,E_{Q_g-q_g})}\frac{2\pi r_h}{f^\prime(r_h)}\sqrt{\frac{2(M-\omega)}{[r_h^2+(Q_g-q_g)^2]^{3/2}}}\left[ \mathcal{W}\mathrm{d}M^\prime-\frac{(Q_g-q_g)}{r_h}\mathrm{d}Q^\prime_g\right].\label{eq:10}
\end{equation}
We get the Hawking temperature:
\begin{equation}
	\beta^\prime=\frac{1}{T^\prime}=-\frac{4\pi r_h}{f^\prime(r_h)}\sqrt{\frac{2(M-\omega)}{[r_h^2+(Q_g-q_g)^2]^{3/2}}}.\label{eq:11}
\end{equation}
Where:
\begin{equation}
	f^\prime(r_h)=-\frac{4(M-\omega)r_h[r_h^2+(Q_g-q_g)^2]-3(M-\omega)r_h^2}{[r_h^2+(Q_g-q_g)^2]^{5/2}}.
\end{equation}
For regular black holes, the modified first law of black hole thermodynamics can be expressed by the following equation:
\begin{equation}
	\mathcal{W}\mathrm{d}M=\frac{T}{4}\mathrm{d}A+V_g\mathrm{d}Q_g.\label{eq:12}
\end{equation}
$\frac{(Q_g-q_g)}{r_h}$ in Eq.~\ref{eq:10} is $V_g$. Put Eq.~\ref{eq:11} and Eq.~\ref{eq:12} into Eq.~\ref{eq:10}, we can get:
\begin{align}
	\mathrm{Im}S=&-\frac{1}{2}\int_{(M,E_{Q_g})}^{(M-\omega,E_{Q_g-q_g})}\frac{1}{T^\prime}\left[\mathcal{W}\mathrm{d}M^\prime-\frac{(Q_g-q_g)}{r_h}\mathrm{d}Q^\prime_g\right]=-\frac{1}{8}\int_{A_i}^{A_f}\mathrm{d}A^\prime\nonumber\\
	=&-\frac{1}{8}(A_f-A_i)=-\frac{1}{2}(S_f-S_i)=-\frac{1}{2}\Delta S_{BH}.
\end{align}
Finally, we get the emission rate of the particle is:
\begin{equation}
	\Gamma\sim e^{-2\mathrm{Im}S}=e^{\Delta S_{BH}}.
\end{equation}
This suggests that the process by which magnetized particles tunnel out of Bardeen's black hole satisfies the unitary principle of quantum mechanics. That is, information is conserved during this process.
\section{Conclusion and Discussion}
\label{sec:3}
We use the Parikh-Wilczek method in our calculation, and finally conclude that the radiation spectrum of the black hole deviates from the pure thermal spectrum, satisfying the unitarity principle. Our results support the conservation of information. The fundamental reason lies in the fact that the tunneling process of particles is reversible within the framework of the Parikh-Wilczek method. In this process, the total entropy of the black hole and the matter and field outside the black hole is conserved, so the information must be conserved.

\section*{Acknowledgements}
The authors would like to thank Prof. Jian Jing and his student LiuBiao Ma and Zheng Wang from the Department of Physics , Beijing University of Chemical Technology for their valuable comments and suggestions during the completion of this manuscript.


\end{document}